\documentclass[preprint,amsmath,superscriptaddress,amssymb,floatfix,nofootinbib]{revtex4}
\usepackage{epsf}
\usepackage{graphicx}
\usepackage{subfigure}
\usepackage{feynmf}
\usepackage{tikz}
\usepackage{epstopdf}
\newcommand{\beq}{\begin{equation}}
\newcommand{\eeq}{\end{equation}}

\begin{document}

%\preprint{~~PITT-PACC-????}

\title{$\Lambda_c^{+}/\Lambda_c^{-}$ and $\Lambda_b^0/\overline{\Lambda}_b^0$ production asymmetry at the LHC from heavy quark recombination}

\def\pitt{Pittsburgh Particle Physics Astrophysics and Cosmology Center (PITT PACC)\\
Department of Physics and Astronomy, University of Pittsburgh, Pittsburgh, PA 15260, USA
\vspace*{.2cm}}

\author{Wai Kin Lai\footnote{Electronic address: wal16@pitt.edu}}
\affiliation{\pitt}

\author{Adam K. Leibovich\footnote{Electronic address: akl2@pitt.edu}}
\affiliation{\pitt}

%\author{A. A. Petrov\footnote{Electronic address: apetrov@wayne.edu}}
%\affiliation{\wsu}\affiliation{\mctp}

\date{\today}

\begin{abstract}
The asymmetries in the forward region production cross section of $\Lambda_c^{+}/\Lambda_c^{-}$ 
and $\Lambda_b^0/\overline{\Lambda}_b^0$ are predicted using the heavy quark recombination mechanism for $pp$
collisions at $7$~TeV and $14$~TeV. 
Using non-perturbative parameters determined from various previous experiments, we find that $A_p(\Lambda_c^{+}/\Lambda_c^{-})\sim 1-2\%$
and $A_p(\Lambda_b^0/\overline{\Lambda}_b^0)\sim 1-3\%$ in the forward region covered by the LHCb experiment. 
\end{abstract}

\maketitle

%==============================================================
% contents
%==============================================================

Asymmetries in productions of heavy hadrons and antihadrons could provide clues to CP violation and physics beyond the standard model. At the LHC, since the initial state $pp$ has a positive baryon number, it is not obvious that productions of heavy hadrons and antihadrons are symmetric even within QCD. In fact, the CERN Intersecting Storage Rings observed an asymmetry in the number of $\Lambda_c^+$ versus $\Lambda_c^-$ \cite{Lockman:1979aj,Chauvat:1987kb}, where production of $\Lambda_c^+$ was favored over $ \Lambda_c^-$ in proton-proton collisions.  On the other hand, the lowest-order QCD cross sections  predict equal numbers of charm and anticharm baryons.    More recently, the CMS detector has measured a small asymmetry in $\Lambda_b$ and $\bar \Lambda_b$ production \cite{Chatrchyan:2012xg}, in particular in the largest rapidity bin.  There have been many models to try to explain this phenomenology, for instance \cite{Andersson:1983ia,Brodsky:1980pb,Brodsky:1981se,Rosner:2012pi,Rosner:2014gta}.
In this paper, we show that the asymmetry can be explained by the QCD-inspired heavy quark recombination mechanism \cite{Braaten:2001bf,Braaten:2001uu,Braaten:2002yt,Braaten:2003vy}, and that the LHCb experiment should see an asymmetry in both $\Lambda_c$ and $\Lambda_b$ production. 

A similar asymmetry can be seen in the meson system.  The asymmetry in $D^{\pm}$ production has been observed in the forward region at LHCb \cite{LHCb:2012fb}, and can also be explained using  the heavy quark recombination mechanism \cite{Lai:2014xya}. The heavy quark recombination mechanism also successfully explained production asymmetries of $D^\pm$ and $\Lambda_c^\pm$ in fixed-target experiments~\cite{Aitala:1996hf,Braaten:2001uu,Braaten:2002yt,Braaten:2003vy}. Therefore, the prediction on production asymmetries of $\Lambda_c^{+}/\Lambda_c^{-}$ and $\Lambda_b^0/\overline{\Lambda}_b^0$ using the heavy quark recombination mechanism should give a sensible estimate for the contribution to the asymmetries from the standard model.

The production asymmetry $A_p$ of a $\Lambda_Q$ $(udQ)$ baryon is defined by
\begin{equation}
A_p=\frac{d\sigma(\Lambda_Q)-d\sigma(\overline{\Lambda}_Q)}{d\sigma(\Lambda_Q)+d\sigma(\overline{\Lambda}_Q)}\,,  \label{eq:asymmetry}
\end{equation}
where $Q$ is either $c$ or $b$. Factorization theorems of perturbative QCD~\cite{Collins:1985gm} state that heavy hadron production cross section can be written in a factorized form. At the LHC, the cross section for producing a $\Lambda_Q$ baryon in a $pp$ collision, at leading order in a $1/p_T$ expansion, is given by
\begin{equation}
d\sigma[pp\rightarrow \Lambda_Q+X]=\sum\limits_{i,j}f_{i/p}\otimes f_{j/p}\otimes d\hat{\sigma}[ij
\rightarrow Q+X]\otimes D_{Q\rightarrow \Lambda_Q}\,,    \label{eq:pQCD}
\end{equation}
where $f_{i/p}$ is the partion distribution function for parton $i$ in the proton, $d\hat{\sigma}
(ij\rightarrow Q+X)$ is the partonic cross section and $D_{Q\rightarrow \Lambda_Q}$ is the 
fragmentation function describing hadronization of a heavy quark $Q$ into a $\Lambda_Q$ baryon. The corresponding equation for 
$\overline{\Lambda}_Q$ is obtained by replacing $D_{Q\rightarrow \Lambda_Q}$ by $D_{\overline{Q}\rightarrow \overline{\Lambda}_Q}$. Charge conjugation symmetry in QCD implies that $D_{Q\rightarrow \Lambda_Q} = D_{\overline{Q}\rightarrow \overline{\Lambda}_Q}$. Thus, perturbative QCD predicts that $A_p=0$, up to corrections suppressed by $1/p_T$.

It should be noted that there are corrections to Eq.~(\ref{eq:pQCD}) that scale as powers of 
$\Lambda_{\rm QCD}/m_Q$ and $\Lambda_{\rm QCD}/p_T$. One should expect non-vanishing power-suppressed contributions to 
$A_p$ at low $p_T$.  A QCD-based model for these power corrections is the heavy quark recombination 
mechanism~\cite{Braaten:2001bf,Braaten:2001uu,Braaten:2002yt,Braaten:2003vy}. In this scenario, a light quark and a heavy quark coming out from the hard scattering process combine to form the final state baryon. This process is of order 
$\Lambda_{\rm QCD} m_Q/p_T^2$ relative to Eq.~(\ref{eq:pQCD}). In what follows, after a brief review of the heavy quark recombination mechanism,\footnote{For a full 
review, please see Refs.~\cite{Braaten:2001bf,Braaten:2001uu,Braaten:2002yt,Braaten:2003vy,Chang:2003ag}.} we calculate $A_p$
due to heavy quark recombination for $\Lambda_c^{+}/\Lambda_c^{-}$ and $\Lambda_b^0/\overline{\Lambda}_b^0$. 

\begin{figure}
\begin{center}
\subfigure[]{
\includegraphics[scale=0.07]{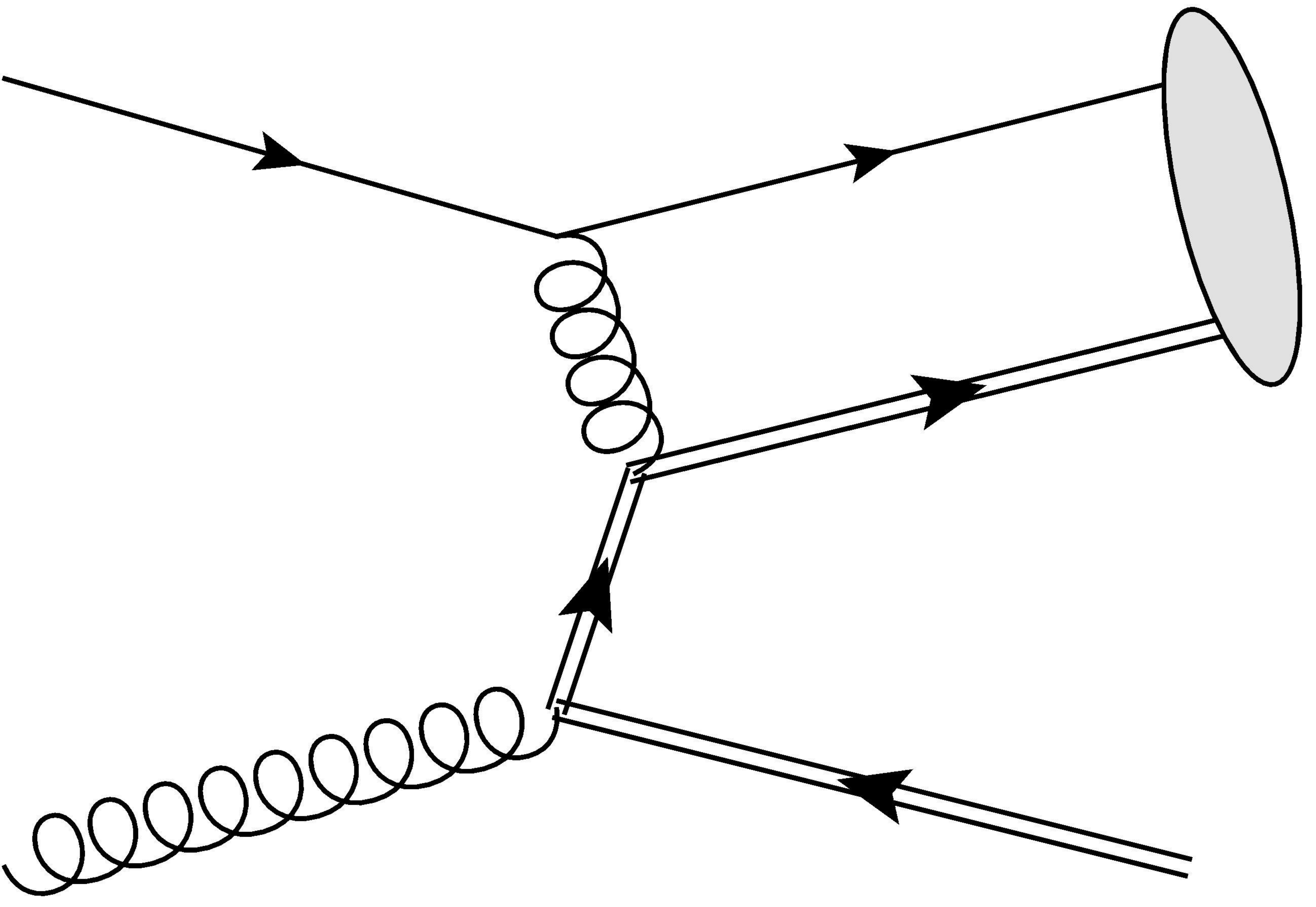}}\label{fg:recomb_qg}
\hskip .5in
\subfigure[]{
\includegraphics[scale=0.06]{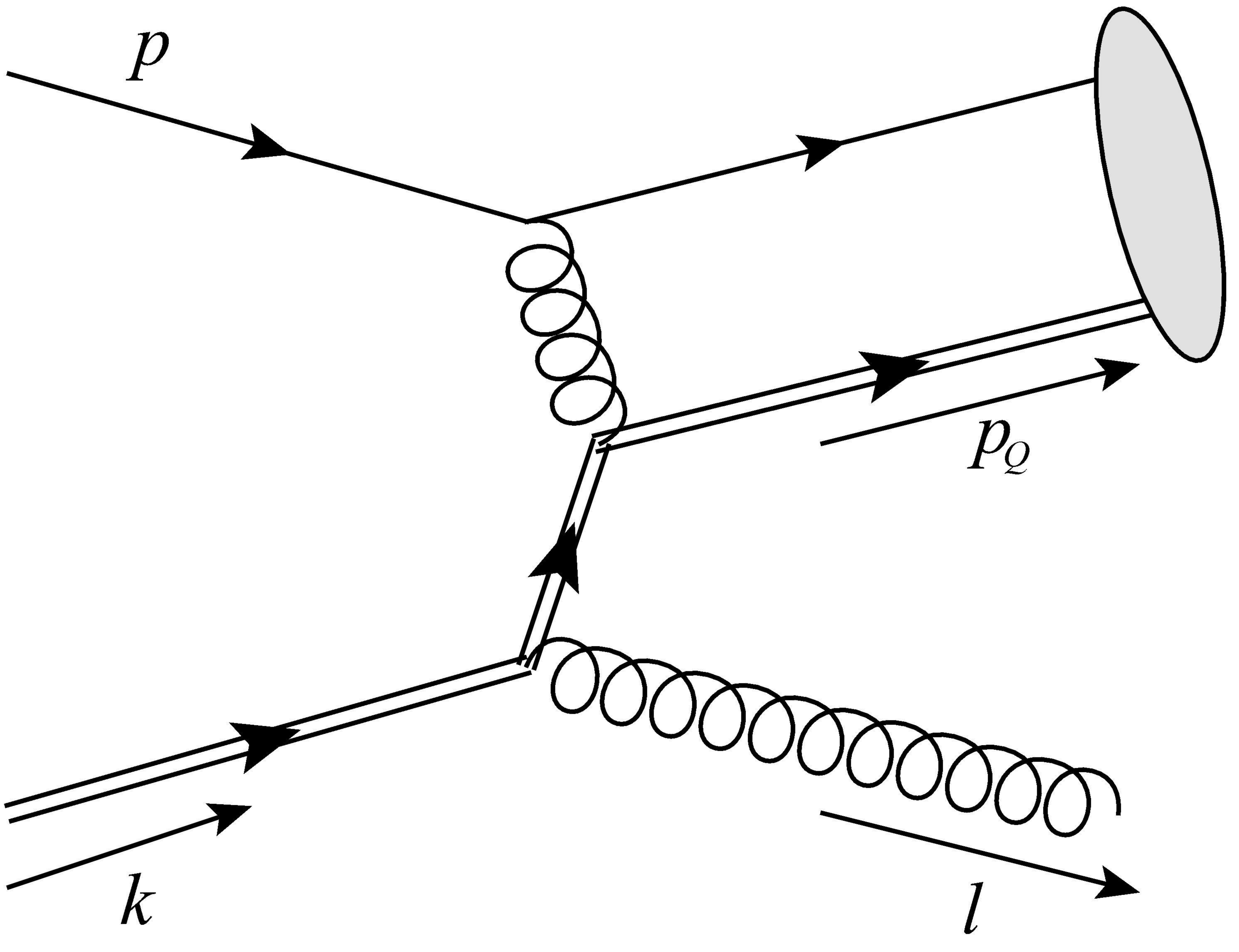}}\label{fg:recomb_Qq}
\vskip .05in  
\caption[]{Diagrams for production of a $\Lambda_Q$ baryon by the heavy-quark recombination mechanism for (a) $qg\rightarrow (Qq)^n+\overline{Q}$
and (b) $Qq\rightarrow (Qq)^n+g$. Each process has five diagrams. Single lines represent light quarks, 
double lines heavy quarks, and the shaded blob the $\Lambda_Q$ baryon. }
\label{fg:recomb}
\end{center}
\end{figure}

Figure \ref{fg:recomb}~(a) and (b) show diagrams for the recombination process with initial state $qg$ and $Qq$ respectively. Their contributions to the cross section are given by
\begin{align}
(a)\phantom{aaaa}d\hat\sigma[\Lambda_Q]&=d\hat{\sigma}[qg\rightarrow (Qq)^n+\overline{Q}]\eta[(Qq)^n\rightarrow \Lambda_Q]\,,\label{eq:recombination_a} \\
(b)\phantom{aaaa}d\hat\sigma[\Lambda_Q]&=d\hat{\sigma}[Qq\rightarrow (Qq)^n+g]\eta[(Qq)^n\rightarrow \Lambda_Q]\,,
\label{eq:recombination_b}
\end{align}
where $(Qq)^n$ indicates that the light quark of flavor $q$ with momentum $\Lambda_{\rm QCD}$ in the $Q$ rest frame is produced in the state $n$, where $n$ labels the color and angular momentum quantum numbers of the quark pair. The cross section is factored into  a 
perturbatively calculable piece $d\hat{\sigma}$ and a nonperturbative factor $\eta[(Qq)^n\rightarrow \Lambda_Q]$ encoding the probability for the quark pair with quantum number $n$ to hadronize into the final state including the $\Lambda_Q$.  
This factorization can be proved within HQET~\cite{Chang:2003ag}. Equations (\ref{eq:recombination_a}) and (\ref{eq:recombination_b}) must then be convoluted with the proton parton distribution functions to get the final hadronic cross section.

The perturbative piece for process (a), was calculated to lowest order in~\cite{Braaten:2003vy}.  
Following the method in \cite{Braaten:2003vy}, we calculate the partonic cross sections for process (b),
$Qq\rightarrow (Qq)^n+g$ (Fig.~\ref{fg:recomb} (b)):
\begin{equation}
\frac{d\sigma}{d\hat{t}}[Qq\rightarrow (Qq)^n+g]=-\frac{2\pi^2\alpha_s^3m_Q^2}{27S^2T}G(n|S,T),
\end{equation}
\begin{align}
G(^1S_0^{(\bar{3})}|S,T)&=-\frac{16S}{T}\left(1-\frac{TU}{S^2}\right)-\frac{m_Q^2}{U}\left(3+\frac{28S}{U}+\frac{16S^2}{U^2}
-\frac{16U^2}{S^2}\right)+\frac{4m_Q^4T}{SU^2}\left(3+\frac{4U}{S}+\frac{8S}{U}\right)\,, \nonumber\\
G(^3S_1^{(\bar{3})}|S,T)&=3G(^1S_0^{(\bar{3})}|S,T)-32\left(\frac{U}{S}-\frac{S^2}{U^2}\right)
-\frac{4m_Q^2}{U}\left(8-\frac{6S}{U}-\frac{16S^2}{U^2}+\frac{13U}{S}+\frac{15U^2}{S^2}\right)\,,\nonumber\\
G(^1S_0^{(6)}|S,T)&=-\frac{4S}{T}\left(2-\frac{5TU}{S^2}\right)-\frac{m_Q^2}{U}\left(27+\frac{14S}{U}+\frac{8S^2}{U^2}-\frac{20U^2}{S^2}\right)+\frac{2m_Q^4T}{SU^2}\left(9+\frac{10U}{S}+\frac{8S}{U}\right)\,, \nonumber\\
G(^3S_1^{(6)}|S,T)&=3G(^1S_0^{(6)}|S,T)-\frac{8S}{T}\left(3+\frac{5TU}{S^2}+\frac{5S}{U}+\frac{2S^2}{U^2}\right)
+\frac{4m_Q^2T}{S^2}\left(27-\frac{S}{U}-\frac{S^2}{U^2}-\frac{8S^3}{U^3}\right)\,,
\end{align}
where we have defined $S = \hat s - m_Q^2 = (k+p)^2 - m_Q^2, T = \hat t = (k - p_Q)^2,$ and $U = \hat u - m_Q^2 = (k - l)^2 - m_Q^2$.

%==========================================================================

%============================================================================

The heavy quark $\overline{Q}$ in the final state of process (a) could fragment into a $\overline{\Lambda}_Q$ baryon. Charge conjugation of this process gives the ``opposite-side recombination":
\begin{align}
(c)\phantom{aaaa}d\hat\sigma[\Lambda_Q]&=\sum\limits_nd\hat{\sigma}[qg\rightarrow (\overline{Q}q)^n+Q]\sum\limits_{\overline{H}_{meson}}\rho[(\overline{Q}q)^n\rightarrow \overline{H}_{meson}]\otimes
D_{Q\rightarrow \Lambda_Q}\,, \label{eq:opp_recombination_meson}\\
(d)\phantom{aaaa}d\hat\sigma[\Lambda_Q]&=\sum\limits_nd\hat{\sigma}[\bar{q}g\rightarrow (\overline{Q}\bar{q})^n+Q]\sum\limits_{\overline{H}_{baryon}}\eta[(\overline{Q}\bar{q})^n\rightarrow \overline{H}_{baryon}]\otimes
D_{Q\rightarrow \Lambda_Q}\
\label{eq:opp_recombination_baryon}
\end{align}
where $H_{meson}$ and $H_{baryon}$ are any heavy meson and heavy baryon respectively. We will assume charge conjugation symmetry and take $\rho[(Q\bar{q})^n\rightarrow H_{meson}]=\rho[(\overline{Q}q)^n\rightarrow \overline{H}_{meson}]$ and $\eta[(Qq)^n\rightarrow H_{baryon}]=\eta[(\overline{Q}\bar{q})^n\rightarrow \overline{H}_{baryon}]$. For simplicity, we will take $H$ to be a low-lying heavy hadron. Thus, for $\Lambda_c$ production we will take $H_{meson}$ to be either $D$ or $D^*$, and $H_{baryon}$ be any baryon from the lowest mass $J^p=\frac{1}{2}^+$ and $\frac{3}{2}^-$ heavy baryon $SU(3)$ flavor multiplets, and similarly for $\Lambda_b$ production. We will also assume $SU(3)$ flavor symmetry.
   
The leading nonperturbative parameters $\rho[(c\bar{q})^n\rightarrow D]$ for $D$ mesons are
\begin{align}
\rho_1&=\rho[c\bar{q}(^1S_0^{(1)})\rightarrow D]\,,
&\tilde{\rho}_1&=\rho[c\bar{q}(^3S_1^{(1)})\rightarrow D]\,, \nonumber \\
\rho_8&=\rho[c\bar{q}(^1S_0^{(8)})\rightarrow D]\,,  
&\tilde{\rho}_8&=\rho[c\bar{q}(^3S_1^{(8)})\rightarrow D]\,. \label{eq:rho_D}
\end{align}
For $D^*$ mesons, we can exploit heavy quark spin symmetry and get
\begin{align}
\rho[c\bar{q}(^1S_0^{(c)})\rightarrow D]&=\rho[c\bar{q}(^3S_1^{(c)})\rightarrow D^*]\,, \nonumber\\
\rho[c\bar{q}(^3S_1^{(c)})\rightarrow D]&=\rho[c\bar{q}(^1S_0^{(c)})\rightarrow D^*]\,. \label{eq:heavy_quark_sym}
\end{align}
Similar relations are also true for $B$ and $B^*$.
The leading nonperturbative parameters $\eta[(Qq)^n\rightarrow \Lambda_Q]$ for $\Lambda_Q$ baryons are
\begin{align}
\eta_3&=\eta[Qq(^1S_0^{(\bar{3})})\rightarrow \Lambda_Q]\,,
&\tilde{\eta}_3&=\eta[Qq(^3S_1^{(\bar{3})})\rightarrow \Lambda_Q]\,, \nonumber \\
\eta_6&=\eta[Qq(^1S_0^{(6)})\rightarrow \Lambda_Q]\,,  
&\tilde{\eta}_6&=\eta[Qq(^3S_1^{(6)})\rightarrow \Lambda_Q]\,. \label{eq:rho_Lambda}
\end{align}
All the $\rho$s and $\eta$s scale as $\Lambda_{QCD}/m_Q$.
Contributions of feeddown from heavier baryons in processes (a) and (b) can be taken into account by using the inclusive parameter $\eta_{inc}$:
\begin{equation}
\eta_{inc}[(Qq)^n\rightarrow\Lambda_Q]=\eta[(Qq)^n\rightarrow\Lambda_Q]+\sum\limits_{H_{baryon}\neq\Lambda_Q}\eta[(Qq)^n\rightarrow H_{baryon}]
B[H_{baryon}\rightarrow\Lambda_Q+X] \label{eq:eta_inc}
\end{equation}
With our choice of possible $\overline{H}_{baryon}$ in the opposite-side recombination, by simple quark counting and the fact that $B[\Sigma_Q\rightarrow\Lambda_Q+X]\approx1$, we have
\begin{equation}
\sum\limits_{\overline{H}_{baryon}}\eta[(\overline{Q}\bar{q})^n\rightarrow \overline{H}_{baryon}]\approx
\frac{3}{2}\eta_{inc}[(Qq)^n\rightarrow\Lambda_Q]\,.
\end{equation}
From~\cite{Lai:2014xya}, a reasonable fit to $D^{\pm}$ asymmetry at LHCb gives $0.055<\rho_1<0.065$, $0.65<\rho_8<0.8$, $0.24<\tilde{\rho}_1<0.3$ and $0.24<\tilde{\rho}_8<0.3$ for $\rho[(c\bar{d})^n\rightarrow D^+]$. Best single-parameter fit to $\Lambda_c^{\pm}$ asymmetry in fixed target experiments gives $\tilde{\eta}_3=0.058$ for $\Lambda_c$~\cite{Braaten:2003vy}. We will take $\eta_1=\tilde{\eta}_1=\tilde{\eta}_3=0$ and
$0.052<\tilde{\eta}_3<0.064$ for $\Lambda_c$. For the $\eta$s for $\Lambda_b$ and $\rho$s for $B$, we simply mutliply the $\Lambda_c$ and $D$ counterparts by the theoretical scaling factor $m_c/m_b$. 
We use MSTW 2008 LO PDFs with $m_c = 1.275$~GeV and $m_b = 4.18$ GeV. The fragmentation function $D_{Q\rightarrow \Lambda_Q}$ is taken as
\begin{equation}
D_{Q\rightarrow \Lambda_Q}(z)=f_{\Lambda_Q}\delta(1-z)\,,  \label{eq:Peterson}
\end{equation}
where $f_{\Lambda_Q}$ is the inclusive fragmentation probability. This form of fragmentation function was found to be better than the Peterson form when fitting to fixed target $\Lambda_c^{+}/\Lambda_c^{-}$ asymmetry data~\cite{Braaten:2003vy}. We take $f_{\Lambda^+_c}=0.101$, which is the average of the values listed in~\cite{Abramowicz:2013eja}. $f_{\Lambda^0_b}$ is taken to be $0.09$ from~\cite{Affolder:1999iq}. We use the LO cross section for the perturbative QCD rate Eq.~(\ref{eq:pQCD}). The factorization scale is set to be $\mu_f=\sqrt{p_T^2+m_Q^2}$.  

\begin{figure}
\begin{center}
\subfigure[]{
      \includegraphics[width=0.5\textwidth,angle=0,clip]{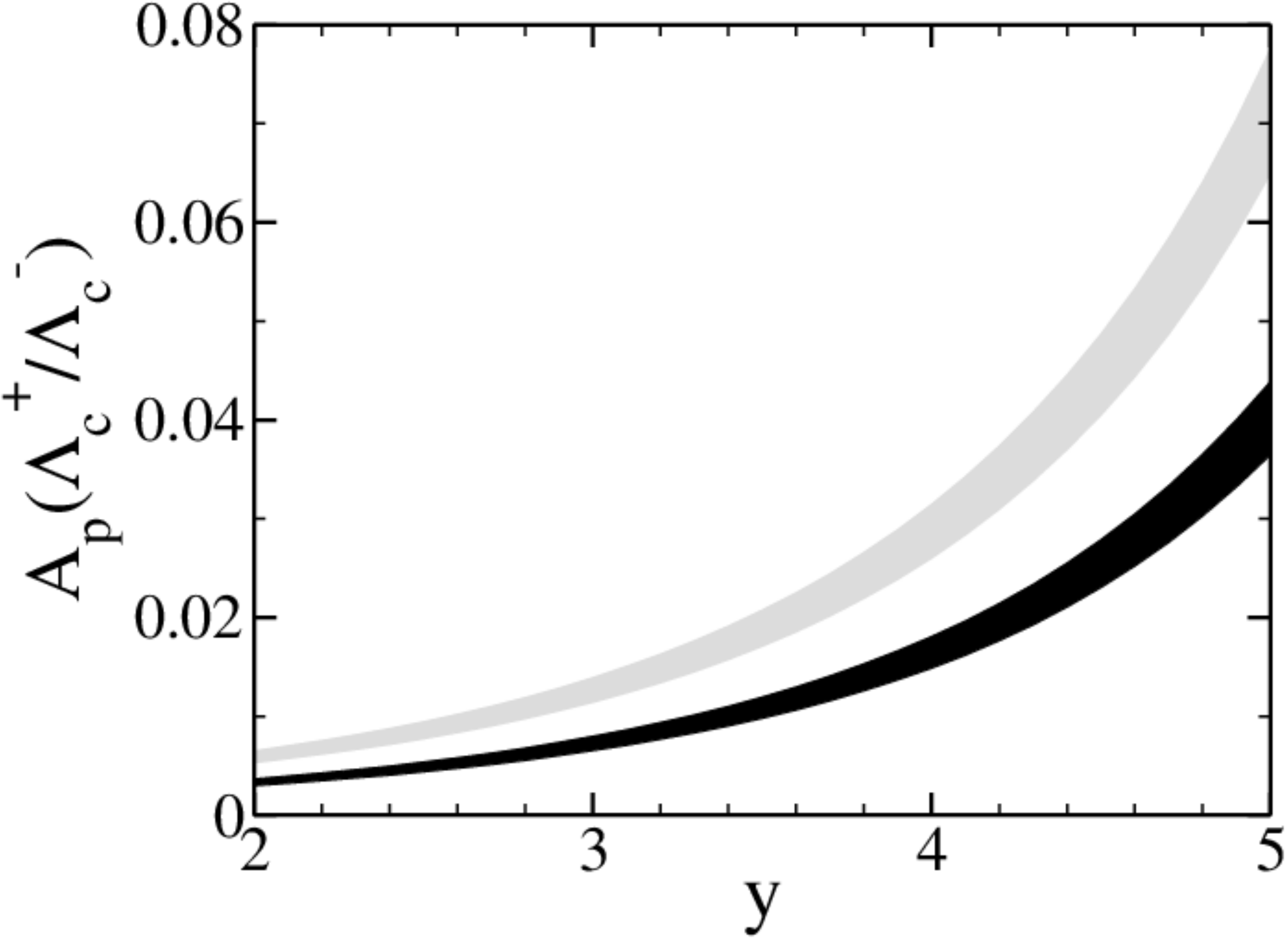}
}
%\hskip .1in
\subfigure[]{
      \includegraphics[width=0.5\textwidth,angle=0,clip]{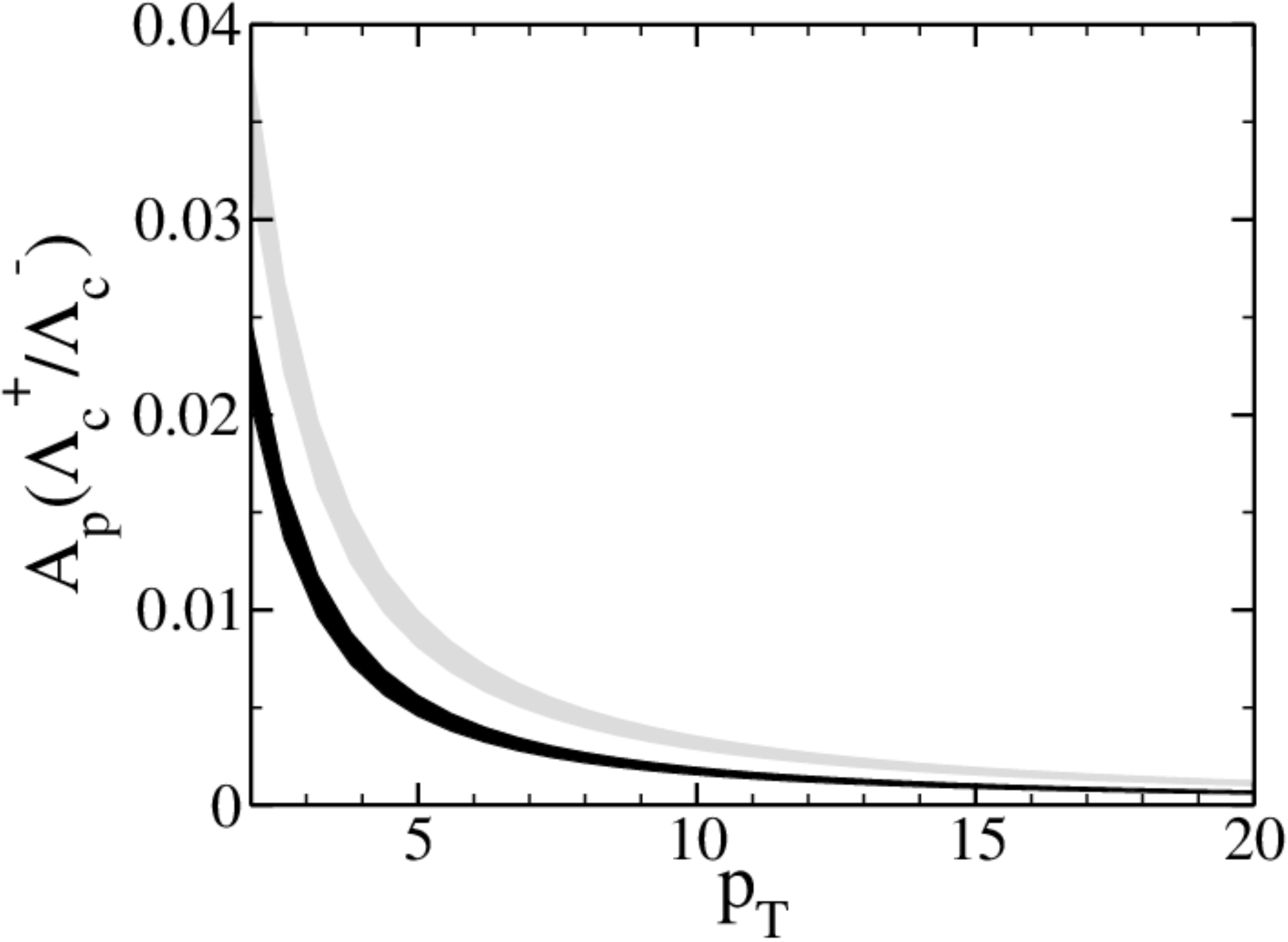}
}

\vskip .1in  
\caption[]{Asymmetry in $\Lambda_c^{+}/\Lambda_c^{-}$ production as a function of (a) rapidity $y$ and (b) transverse momentum $p_T$ in the kinematic region $2<y<5$ and $2{\rm{~GeV}}<p_T<20{\rm{~GeV}}$ in $7$~TeV (grey band) and $14$~TeV (black band) $pp$ collisions.}
\label{fg:Ap_c}
\end{center}
\end{figure}

\begin{figure}

\begin{center}
\subfigure[]{
      \includegraphics[width=0.5\textwidth,angle=0,clip]{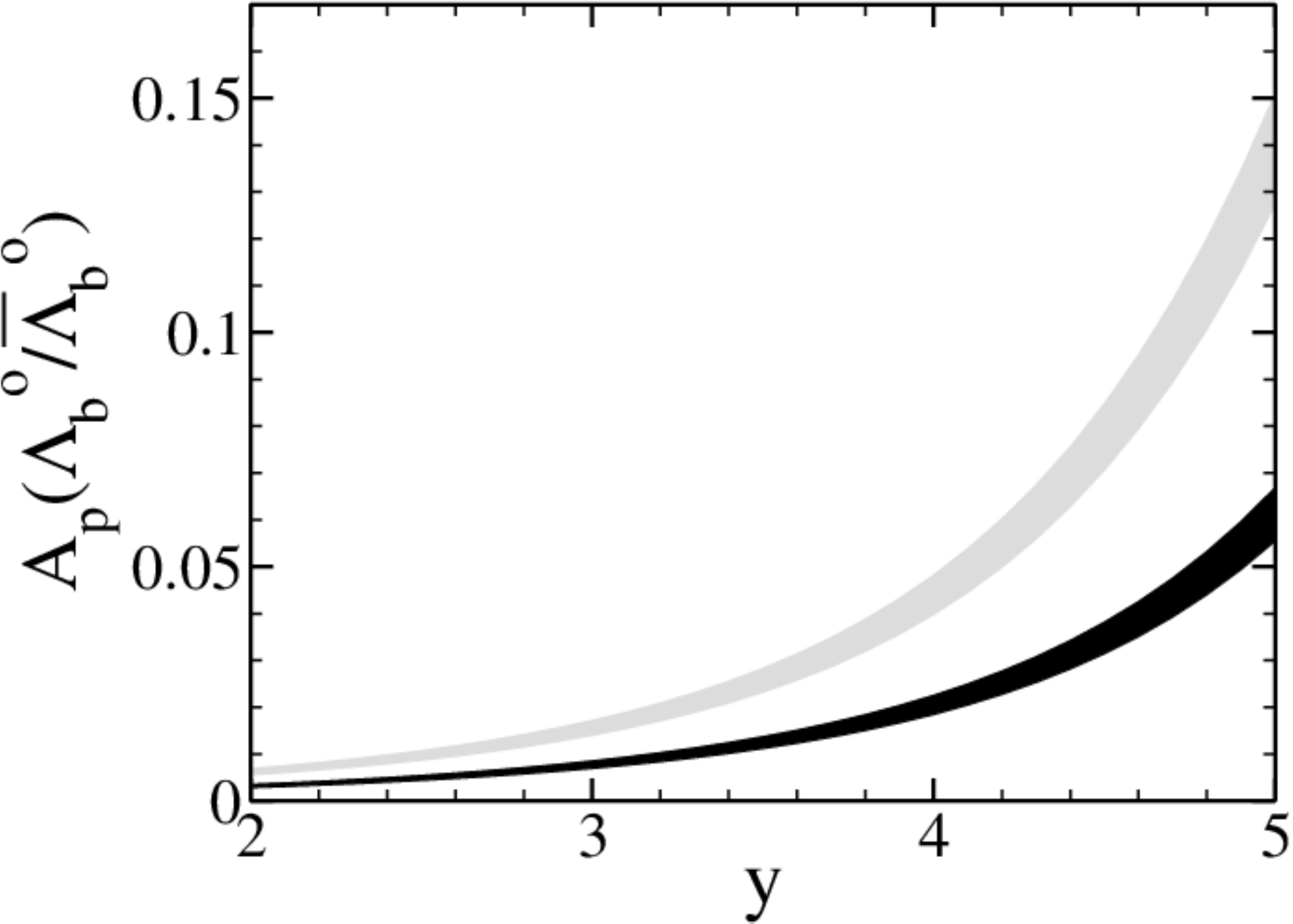}
}

%\hskip .05in
\subfigure[]{
      \includegraphics[width=0.5\textwidth,angle=0,clip]{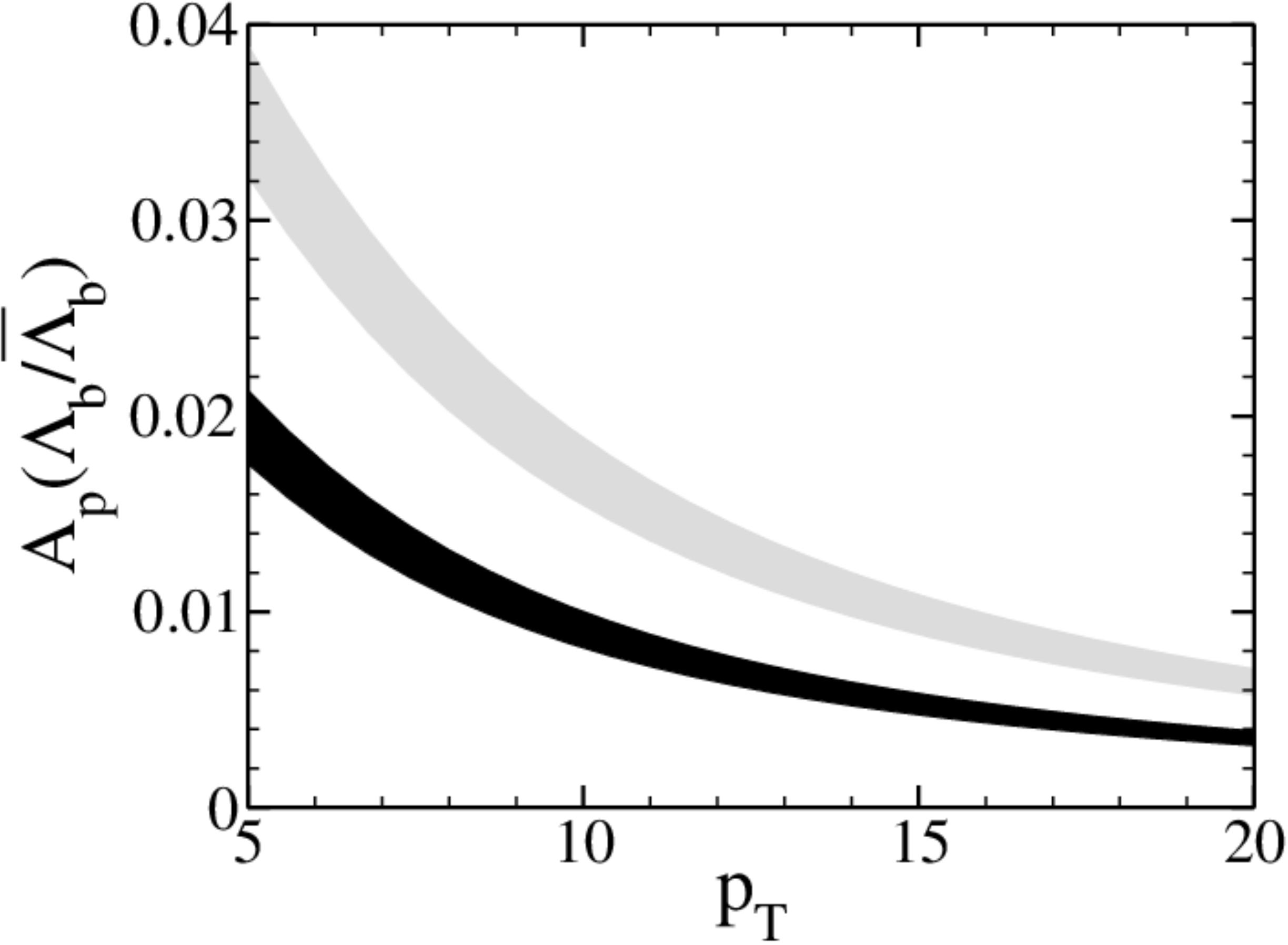}
}

\vskip .1in  
\caption[]{Asymmetry in $\Lambda_b^0/\overline{\Lambda}_b^0$ production as a function of (a) rapidity $y$ and (b) transverse momentum $p_T$ in the kinematic region $2<y<5$ and $2{\rm{~GeV}}<p_T<20{\rm{~GeV}}$ in $7$~TeV (grey band) and $14$~TeV (black band) $pp$ collisions.}
\label{fg:Ap_b}
\end{center}
\end{figure}

For $\Lambda_c^{+}/\Lambda_c^{-}$ productions, the kinematic region is taken to be $2<y<5$ and $2{\rm{~GeV}}<p_T<20{\rm{~GeV}}$.  We find that $2.0\%<A_p(\Lambda_c^{+}/\Lambda_c^{-})<2.4\%$ for $\sqrt{s}=7$~TeV and $1.2\%<A_p(\Lambda_c^{+}/\Lambda_c^{-})<1.5\%$ for $\sqrt{s}=14$~TeV. For $\Lambda_b^0/\overline{\Lambda}_b^0$ productions, the kinematic region is taken to be $2<y<5$ and $5{\rm{~GeV}}<p_T<20{\rm{~GeV}}$. We find that 
$2.2\%<A_p(\Lambda_b^0/\overline{\Lambda}_b^0)<2.6\%$ for $\sqrt{s}=7$~TeV and $1.1\%<A_p(\Lambda_b^0/\overline{\Lambda}_b^0)<1.4\%$ for $\sqrt{s}=14$~TeV. Figures \ref{fg:Ap_c} and \ref{fg:Ap_b} show the rapidity and transverse momentum distributions of $A_p$ for $\Lambda_c^{+}/\Lambda_c^{-}$ and $\Lambda_b^0/\overline{\Lambda}_b^0$ respectively. The asymmetry is significant in the high rapidity and low rapidity ends ($\sim 2-15\%$).  

In summary, we have use the heavy quark recombination mechanism to calculate the production asymmetries for $\Lambda_c^{+}/\Lambda_c^{-}$ and $\Lambda_b^0/\overline{\Lambda}_b^0$ at the LHCb experiment.  The heavy quark recombination mechanism is a correction of order $\Lambda_{\rm QCD} m_Q/p_\perp^2$ to the leading order QCD prediction.  Therefore, we expect large effects at small $p_\perp$ and/or large rapidity.  Our calculation confirms this expectation, where the differential distributions are significant at the high-rapidity and low-$p_T$ ends ($\sim 2-15\%$).  The integrated asymmetries in the LHCb region are of the order of $\sim1-3\%$ and should be measurable.

%

%===============================================================
% acknowledgements
%===============================================================
\section*{Acknowledgements}
AKL and WKL are supported in part by the National Science Foundation under Grant No. PHY-1212635.

%===============================================================
% references
%===============================================================

\end{document}